# A deep neural network for positioning and inter-crystal scatter identification in multiplexed PET detectors

Francisco E. Enríquez-Mier-y-Terán[a,b,*], Luping Zhou[c], Steven R. Meikle[b,d], Andre Z. Kyme[a,b]

[a]School of Biomedical Engineering, Faculty of Engineering, The University of Sydney, Sydney, NSW 2006, Australia
[b]Brain and Mind Centre, The University of Sydney, Sydney, NSW 2050, Australia
[c]School of Electrical and Information Engineering, The University of Sydney, Sydney, NSW 2008, Australia
[d]Faculty of Medicine and Health, The University of Sydney, Sydney, NSW 2050, Australia

*Corresponding author.
E-mail address: fenr7890@uni.sydney.edu.au (F.E. Enríquez-Mier-y-Terán)

## Abstract

Objective:
High-resolution PET relies on the accurate positioning of annihilation photons impinging the crystal array. However, conventional positioning algorithms in light-sharing PET detectors are often limited due to edge effects and/or the absence of additional information for identifying and correcting scattering within the crystal array (known as inter-crystal scattering). This study explores the feasibility of deep neural network techniques for more precise event positioning in finely segmented and highly multiplexed PET detectors with light-sharing.

Approach:
Initially, a GATE simulation was used to study the spatial and statistical properties of inter-crystal scatter (ICS) events in finely segmented LYSO PET detectors. Next, a deep neural network (DNN) for crystal localisation was designed, trained and tested with light distributions of photoelectric (P) and Compton + photoelectric (CP) events simulated using optical GATE and an analytical method to speed up data generation. Using the statistical properties of ICS events, an energy-guided positioning algorithm was then built into the DNN. The positioning algorithm enables selection of the unique or first crystal of interaction in P and CP events, respectively. Performance of the DNN was compared with Anger logic using light distributions from simulated 511-keV point sources placed at different locations around a single PET detector module.

Main results: The fraction of events forward and backward scattered in the LYSO detector was 0.54 and 0.46, respectively, whereas naïve application of the Klein-Nishina formulation predicts 70% forward scatter. Despite coarse photodetector data due to signal multiplexing, the DNN demonstrated a crystal classification accuracy of 90% for P events and 82% for CP events. For crystal positioning, the DNN outperformed Anger logic by at least 34% and 14% for P and CP events, respectively. Further improvement is somewhat constrained by the physics – specifically, the ratio of backward to forward scattering of gamma rays within the crystal array being close to 1. This prevents selecting the first crystal of interaction in CP events with a high degree of certainty.
<u>Significance:</u> Light sharing and multiplexed PET detectors are common in high-resolution PET, yet their traditional positioning algorithms often underperform due to edge effects and/or the difficulty in correcting ICS events. Our study indicates that DNN-based event positioning has the potential to enhance 2D coincidence event positioning accuracy by nearly a factor of 2 compared to Anger logic. However, further improvements are difficult to foresee without additional information such as event timing.

## Introduction

High-resolution positron emission tomography (PET) requires accurate positioning of annihilation photons in the transverse plane and in the depth direction (referred to as depth of interaction, DOI). In practice, event positioning accuracy is highly dependent on the detector design, including the crystal geometry, photodetector configuration, and readout electronics.

In PET, typically two crystal configurations are used: monolithic blocks or pixelated arrays. Positioning in monolithic blocks is based on light spreading across photodetectors and extensive calibrations of light fields (Gonzalez-Montoro et al 2021). Although monolithic detectors are cost-effective compared to their pixelated counterparts, the highest positioning resolution is obtained for thin (≤10 mm) scintillator blocks (Gonzalez et al 2018), resulting in an inherent trade-off with sensitivity.

For pixelated arrays the spatial resolution depends primarily on the transverse crystal size, while sensitivity – as for monolithic blocks – depends on the crystal thickness. Positioning in pixelated detectors further depends on the crystal to photodetector ratio. For 1:1 coupling (i.e., where each individual crystal is coupled to its own unique photodetector), positioning is performed by assigning the event to the crystal that caused the photodetector to fire. However, for high resolution applications, such as small animal PET, where finely pixelated and tightly packed crystal arrays are needed (Stickel et al 2007, Yang et al 2016, Yamamoto et al 2016), 1:1 coupling is impractical due to space and cost constraints. Instead, such systems usually rely on light-sharing schemes in which groups of crystals are coupled to the same photodetector (typically via a light guide) (Fig. 1a), and event positioning is based on centre-of-mass calculations.

Although light-sharing schemes can achieve similar performance to 1:1 coupling (Casey and Nutt 1986, St. James and Thompson 2006), there are two common drawbacks: poor performance at the edges of the crystal array due to insufficient light sharing among neighbouring photodetectors; and the inability to identify and correctly position events within the same detector array due to inter-crystal scattering (ICS). As shown in Fig. 1b-c, the multiple interactions of annihilation photons within the crystal array in ICS events can introduce event positioning error that propagates to the reconstruction and, ultimately, compromises the spatial resolution of the system (Shao et al 1996, Zhang et al 2019).

To address edge effects, different hardware-based solutions have been proposed, including the addition of slits into the light guides (Pichler et al 2004, Song et al 2010) and using more complex detector designs (Liu et al 2023). The degradation caused by ICS can be handled by rejecting these events (e.g., Kang et al 2021), however this is at the expense of system sensitivity. Instead, approaches based on ICS identification and correction are preferable. Improved positioning of ICS events based on maximum likelihood (Shinohara et al 2014) and convex constrained optimisation (Lee et al 2018) are feasible but offer only minor improvements (compared to Anger logic), even at low light sharing ratios.

More recently, positioning algorithms based on deep neural networks (DNNs) have been reported for both monolithic PET detectors (Iborra et al 2019, Sanaat and Zaidi 2020, Decuyper et al 2021, Jaliparthi et al 2021) and light-sharing pixelated PET detectors (LaBella et al 2020, Lee and Lee 2021, Lee and Lee 2023, Petersen et al 2024). In the case of light-sharing pixelated detectors, convolutional neural networks (CNNs) can improve event positioning through edge crystal identification (LaBella et al 2020), DOI estimation (Petersen et al 2024) and ICS correction (Lee and Lee 2021, Lee and Lee 2023, Petersen et al 2024). For example, Lee and Lee (2023) reported ~40% accuracy in selecting the first crystal of interaction in ICS events and a resulting volumetric resolution improvement of 47-64% in a detector with high light sharing ratio. However, all of the prior work assumes that each individual photodetector signal is accessible for analysis, which is not always the case. Signal multiplexing of the electronic readout is often used to further simplify the electronic configuration and overall cost (Kyme et al 2017, Yang et al 2019, Park et al 2022). It is not clear how NN design and implementation should change in this regime in order to provide superior event positioning compared to traditional positioning methods (i.e., Anger logic).

The aim of this Monte Carlo simulation study was to investigate whether DNN-based algorithms can outperform Anger logic for event positioning in light sharing and multiplexed PET detectors. We start by characterising ICS events in finely segmented LYSO detectors. We then introduce a DNN-based pipeline designed for 2D event positioning and multiplexed photodetector data. Proof-of-concept of the DNN-based positioning algorithm is demonstrated using optical Monte Carlo data from spatially distinct sources and compared to that of Anger logic. While this work does not encompass DOI estimation, we describe how the method can be adapted to include DOI within the pipeline.

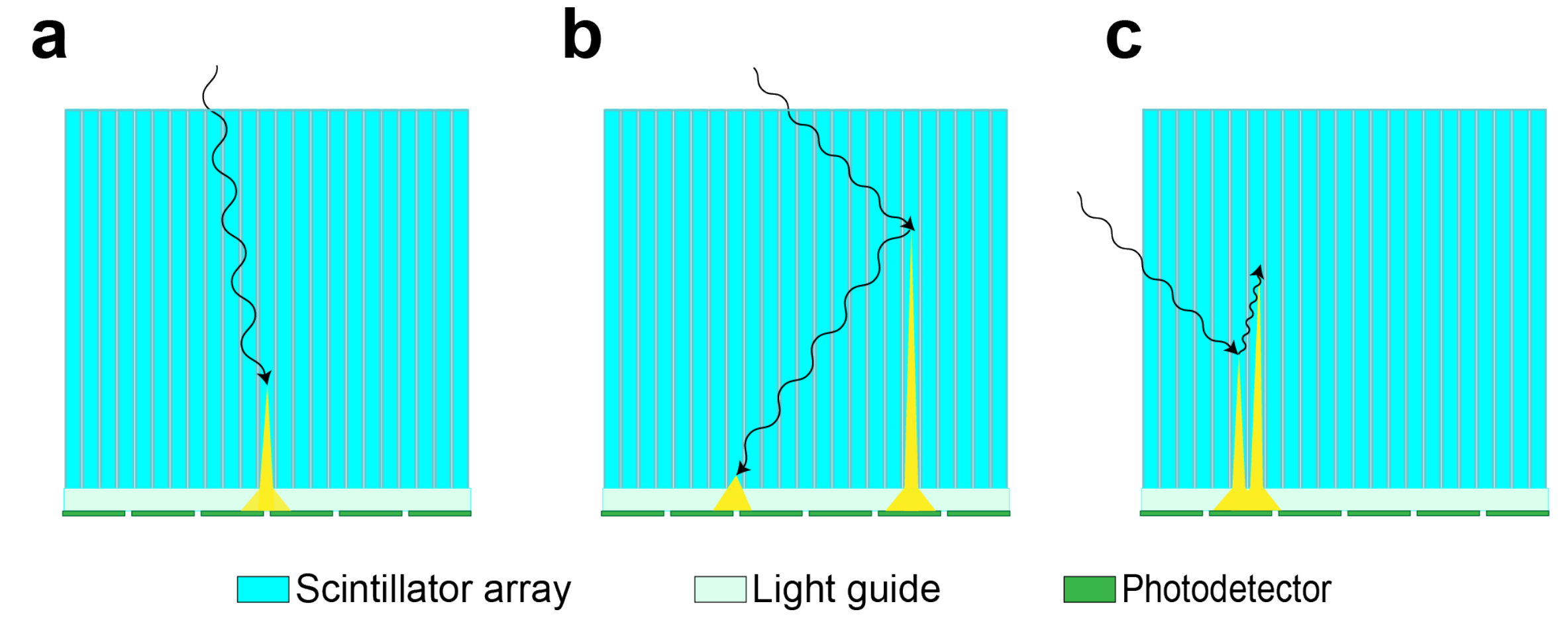

**Figure 1**. Light spreading in light-sharing PET detectors. (**a**) Optical photons emitted from a photoelectric interaction in the scintillation crystal array are detected by two of six photodetectors at one end of the array. Inter-crystal scatter (ICS) events in the forward (**b**) and backward (**c**) directions produce optical photons at multiple locations in the crystal array and the event position is typically estimated based on the centre-of-mass of the total light distribution.

## 2. Methods

### 2.1. ICS in Finely Segmented LYSO crystals

To characterise the spatial and statistical distribution of ICS events in finely segmented LYSO PET detectors, we used Geant4 Application for Tomographic Emission (GATE) v8.2 (Sarrut et al 2021) to simulate the high resolution open-field mouse brain PET scanner (Enriquez-Mier-Y-Teran et al 2021). The scanner is composed of four detector panels with each panel supporting 4 × 4 detector blocks and each block comprising 23 × 23 LYSO crystals. Individual crystals are 0.785 × 0.785 × 20 $mm^3$ (pitch = 0.85 mm), forming a total block size of 19.6 × 19.6 × 20 $mm^3$. Two photodetector arrays, each consisting of a 6 × 6 array of silicon photomultipliers (SiPMs) (3.16 × 3.16 $mm^2$ TSV J-Series, ONSEMI, USA) (pitch = 3.3 mm), read out the scintillation light from both ends of each detector block for DOI estimation.

Events were generated using a point source positioned at the centre of the FOV and emitting back-to-back 511 keV gamma rays. We focused on events characterised by a photoelectric interaction as the sole interaction (a 'P' event), or as the last interaction after a series of Compton scatter interactions ('CP', 'CCP', '3CP', etc events). In a light-sharing PET detector, all of these events would fall within the photopeak window since they finish with a P interaction (i.e. the total energy of the original annihilation photon is eventually deposited in the crystal). We simulated ~60 million events, for each one recording the number of scatter interactions (prior to the photoelectric interaction) within the same detector block, and the ID of the crystals in which the events interacted. Interactions happening within the same crystal were ignored since they only impact DOI positioning accuracy without impacting the 2D positioning of the annihilation photons. Thus, a CP event occurring within the same crystal was treated as a P event for the purposes of this study. CP events were later categorised into *neighbouring* CP events (nCP) and *distant* CP events (dCP) based on the crystal index difference – what we refer to as the crystal window – between the scatter and photoelectric interaction (Fig. 2a). Four crystal windows were considered with index differences ranging from 0 to 3 crystals. For each crystal window we analysed the backward and forward scattering ratios, which in turn guided the DNN-based positioning algorithm in selecting the crystal of first interaction (Section 2.3).

### 2.2. DNN-based Positioning

In this work we used optical GATE simulations and an efficient analytical method to generate the training dataset. We focused exclusively on P and CP events, with the rationale for this choice becoming evident in Section 3.1.

#### 2.2.1. Optical Simulations

Optical simulations were performed using the single PET detector described in section 2.1. Teflon wrapping of each LYSO crystal was modelled using the 'PolishedTeflon_LUT' surface (Roncali et al 2017, Stockhoff et al 2017). Crystal to photodetector coupling was via 1.2 mm thick PMMA light guides (refractive index 1.5) with polished surfaces (UNIFIED model, Levin and Moisan 1996). Photon detection within the active areas of the photodetector array was implemented using the 'Detector_LUT' surface with an efficiency of 1. The 35% SiPM photon detection efficiency (PDE) was modelled by randomly deleting 65% of photons from each SiPM in a post-processing step. Further information on the optical properties of LYSO crystals can be found in (Van Der Laan et al 2010, Stockhoff et al 2019).

We simulated a 511-keV pencil beam gamma source moving perpendicular to the crystal front face, and collected 40,000 photoelectric events for each of 66 neighbouring crystals (corresponding to 1/8 of the array) (Fig. 2b). Optical photons produced from these photoelectric events were collected to simulate the two (i.e. front and back) $6 \times 6$ SiPM array light distributions, based on the number of photons reaching each SiPM. To include the light contribution from the remainder of the array we relied on symmetry, mirroring the light distributions along the $X$ and $Y$ axes. Since we ignored DOI for this study, the front and back $6 \times 6$ light distributions were combined to enhance the dataset statistics. To account for the electronic signal multiplexing of the photodetector data in our small animal system (Kyme et al 2017), the $6 \times 6$ light distributions were compressed into a $1 \times 6$ row and a $1 \times 6$ column distribution by summing the values along rows and columns, respectively. Finally, we calculated the mean and variance of the row and column light distributions for each crystal in the array. Given that each crystal has a unique combination of row and column light distributions, we refer to them as 'basis distributions'.

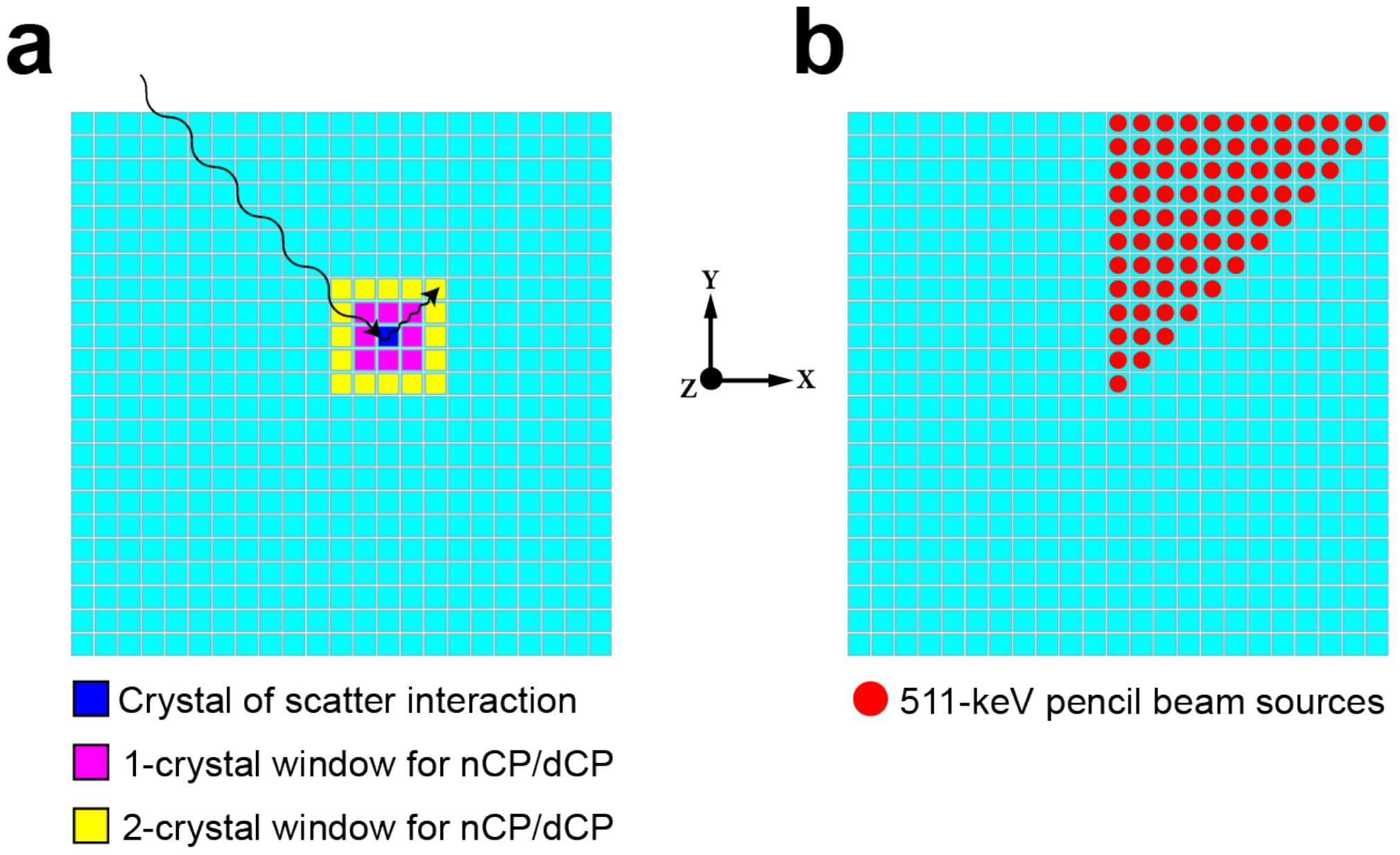

**Figure 2**. (**a**) A "CP event", in which an incoming 511-keV photon scatters inside a crystal (dark blue) before undergoing a photoelectric interaction. CP events were divided into two groups: neighbouring (nCP) and distant (dCP). Photoelectric interactions of the scattered photon within a defined window of either 1 crystal (magenta region) or 2 crystals (yellow region) from the crystal of first interaction are considered nCP events; if outside this window they are treated as dCP events. For example, using a 1-crystal window the scattered photon interacting in the yellow zone corresponds to a dCP event whereas using a 2-crystal window this still would be considered a nCP event. (**b**) To determine basis distributions, 1/8 of the crystal array was irradiated with monoenergetic (511 keV) pencil beam sources (red circles) impinging perpendicular to the crystal front face. Only the light distributions from photoelectric events were recorded.

2.2.2. CP Events as Weighted Combinations of Basis Distributions

With the basis distributions established, we simulated the row/column distributions from CP events using an energy-weighted combination of the basis distributions corresponding to the crystals involved in the CP event:

$$CP^{row} = e_a \times row_a + e_b \times row_b \quad (1)$$

$$CP^{col} = e_a \times col_a + e_b \times col_b \quad (2)$$

where $CP^{row/col}$ is the row/column light distribution of a CP event involving crystals $a$ and $b$; $row_a/col_a$ and $row_b/col_b$ are the row/column basis distributions of crystals $a$ and $b$, respectively; and $e_a$ and $e_b$ are the energies deposited in each crystal, with $e_b = (511 - e_a)$ keV.

Using the basis distributions and their variance, together with the crystal IDs obtained from CP events in section 2.1, we generated a dataset comprising approximately 127 million distinct CP row-column light distributions for training the NN. The CP light distributions were randomly combined with approximately 19 million light distributions from photoelectric events. Finally, the light distributions were made independent of energy by normalising to the area under the curve. This enabled handling of ICS events of diverse energies (e.g., CC events at or below the energy window) and with different SiPM overvoltages. Post-normalisation, the sum of the energy weights is constrained to unity (i.e. $e_a + e_b = 1$).

2.2.3. Neural Network Architecture and Training Method

The DNN was implemented using Keras and TensorFlow-GPU v2.6.2 and an NVIDIA GeForce RTX 2060 GPU. The architecture consisted of 12 input neurons (6 for the row light distributions and 6 for the column light distributions), 3 fully-connected hidden layers each with 512 neurons, and 1 partially-connected hidden layer with 936 neurons and 93 output neurons (Fig. 3c). All hidden layers were activated with rectified linear unit (ReLU) functions. Four groups of 23 output neurons each were activated using softmax functions. These neurons represent the row and column crystal indices ($c_a^{row}, c_a^{col}, c_b^{row}, c_b^{col}$), $a, b$ ∈ {0-22}, and predict the likelihood (0-1) of a crystal being involved in a P and/or CP event. For each group, the crystal index with the highest likelihood was chosen as the crystal of interaction (Fig. 3c). The 2D crystal locations were then:

$$\boldsymbol{c}_a = \left(\arg max\,(\boldsymbol{c}_a^{row}), \arg max\,(\boldsymbol{c}_a^{col})\right) \tag{3}$$

$$\boldsymbol{c}_b = \left(\arg max\,(\boldsymbol{c}_b^{row}), \arg max\,\left(\boldsymbol{c}_b^{col}\right)\right) \tag{4}$$

The last output neuron was activated with a sigmoid function to predict the energy deposition in crystal $a$ ($0 \leq e_a \leq 1$). For P events $e_a$ is set equal to 1.

The loss function comprised five terms:

$$Loss = \left\|e_a^{pred} - e_a^{true}\right\|^2 + CE(\boldsymbol{c}_a^{row}) + CE(\boldsymbol{c}_b^{row}) + CE(\boldsymbol{c}_a^{col}) + CE\left(\boldsymbol{c}_b^{col}\right) \tag{5}$$

The first term regresses the error between the true ($e_a^{true}$) and predicted energy ($e_a^{pred}$) and the remaining four terms are the categorical cross-entropy ($CE$) for each row and column crystal index. Combined, the loss minimises the energy and crystal index discrepancy between true and predicted. Training consisted of a 'warm-up' (20 epochs) involving only the CE terms of the loss function, followed by an alternating-loss approach in which we switched between the energy regression term and the four crystal index classification terms in 5-epoch steps for a total of 600 epochs. The learning rate was set to 0.001 and mini batches were used to avoid saturating the GPU. To avoid overfitting the network a dropout of 0.5 was used in all hidden layers during network training.

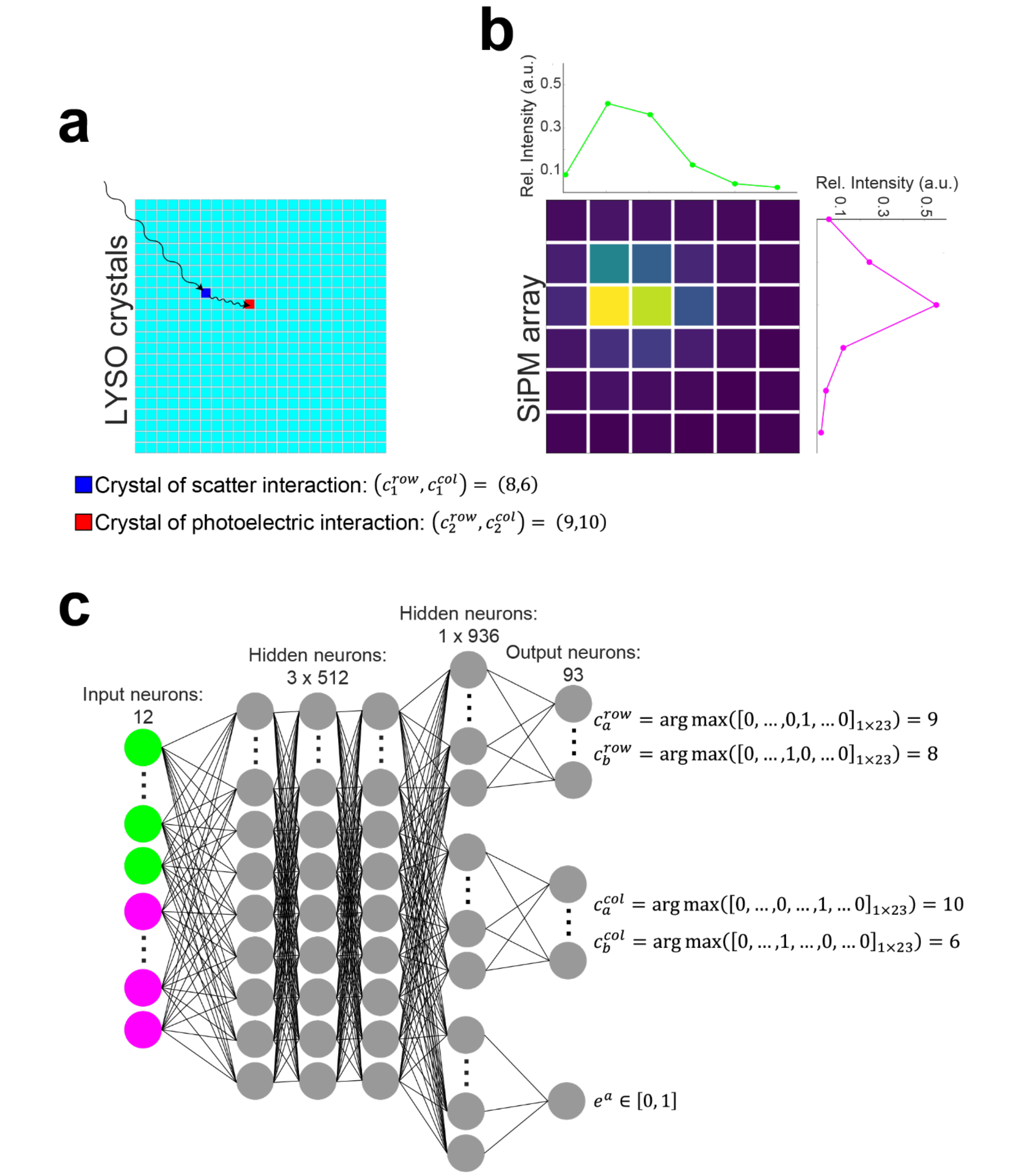

**Figure 3**. DNN-based crystal prediction. (**a**) A 511-keV gamma ray scatters inside crystal $\boldsymbol{c_1}$ (dark blue), with row-column indices (8,6), and then undergoes a P interaction inside crystal $\boldsymbol{c_2}$ (red), at indices (9,10). (**b**) Light produced at both interactions is recorded on the SiPM array and the row-column signal reduction algorithm is applied giving rise to the light distributions shown in magenta and green, respectively. (**c**) Normalised row (magenta) and column (green) light distributions are fed into the neural network. The row and column crystal indices with highest likelihood are selected as the crystals of interaction.

2.3. DNN-based Positioning Algorithm

2.3.1. Algorithm Implementation

Since no timing information was considered during network training, we did not assume that the neural network could discover the order of CP interactions – that is, whether $\boldsymbol{c}_a$ or $\boldsymbol{c}_b$ was the first crystal of interaction ($\boldsymbol{c}_1$). To select $\boldsymbol{c}_1$, a guided discrimination was implemented. While it may be tempting to assume that most CP events involve forward scattering according to the Klein-Nishina formulation (Klein and Nishina 1929), results in section 3.1 demonstrate that this assumption does not hold for small crystals - and indeed penalises the network positioning performance if it is imposed. Instead, we classified the network outputs into three categories - P, nCP, and dCP events - prior to selecting the first crystal of interaction. To categorize the events, we relied on the difference between the two predicted crystals (i.e., $\boldsymbol{c}_a - \boldsymbol{c}_b$). P events exhibited a crystal difference of 0, while nCP and dCP events were distinguished using a 1-crystal window, as outlined in section 2.1. To position P events, either of the two crystal indices could be chosen since they were the same. To position CP events, we based the choice on Compton kinematics. Here we assumed that all nCP events involved backward scattering, selecting $\boldsymbol{c}_1$ as the one with the highest energy deposition (i.e., $max(e_a,\ e_b$ = 1 - $e_a$)). Conversely, we assumed that all dCP events involved forward scattering and chose $\boldsymbol{c}_1$ as the one with the lowest energy deposition (i.e., $min(e_a,\ e_b)$).

2.3.2 Performance Comparison

To assess the DNN-based positioning performance, we simulated a single PET detector in GATE (see Section 2.1) and three point sources emitting isotropic 511-keV gamma rays. The point sources were positioned 3 cm from the detector front face along the $Z$-axis and at different locations in the $XY$-plane (Fig. 4). Point source 3 was offset by 2 cm to the side of the detector so that gamma rays impinged on the detector obliquely. Using the DNN output and our scatter assumptions, the position of P and CP events was estimated and compared to that of a conventional approach using Anger logic. For Anger logic, the detector readout circuitry, based on a resistor ladder approach (Kyme et al 2017), was simulated, excluding the amplification stages. The row-column light distributions were reduced to four signals ($X^+$, $X^-$, $Y^+$, and $Y^-$), which were then combined to calculate the $x$ and $y$ position values. A segmented crystal map was generated using the basis distributions from section 2.2.1 and the watershed method (Vincent and Soille 1991). The map was used to translate the $x$ and $y$ positions into row and column crystal indices ∈ {0-22}. Performance was quantified based on *L1* distance ($\|\|_1$) comparisons between the ground-truth positioning data, obtained from the simulation, and the crystal predictions generated by the DNN approach and the Anger logic algorithm.

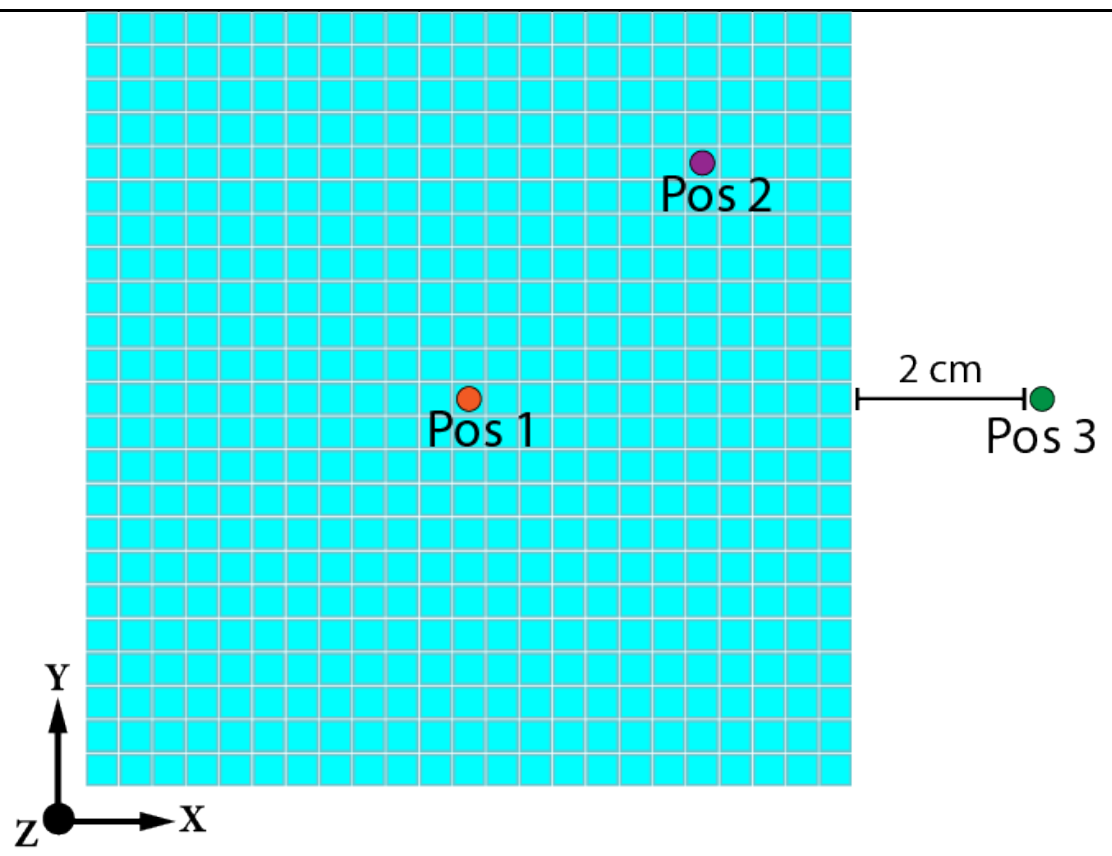

**Figure 4**. Three point sources were simulated to test the performance of the DNN positioning approach. In all cases, the distance along the *Z* axis between the crystal array front face and the point sources was 3 cm.

## Results

3.1. ICS Statistical Distribution

Figure 5a displays the statistical distribution of P and ICS events terminating in a photoelectric interaction. Out of 60 million recorded events, 55% involved photoelectric interaction without scattering in the crystal array. Events characterized by a single scatter preceding photoelectric absorption accounted for 34%. Multi-scatter ICS events, denoted as 'other' in the plot, constituted only 11% of the total events. The distribution of multi-scatter ICS events is depicted in figure 5b, which indicates that the vast majority (~83%) were CCP. Figures 6a-d illustrate the forward and backward scatter ratios of CP events, categorised as neighbouring (nCP) and distant (dCP) events. In Figure 6a, no crystal window was considered, and all events were treated as dCP events, resulting in forward and backward scatter ratios of 0.54 and 0.46, respectively. With a 1-crystal window (Fig. 6b), the forward and backward ratios for dCP events were 0.60 and 0.40, while for nCP events these ratios were 0.41 and 0.59, respectively. As the window size increased to 2 and 3 crystals (Figs. 6c and 6d), there was a shift in favour of forward scattering for dCP events and a shift in the opposite direction, resulting in a flattening of the ratio, for nCP events.

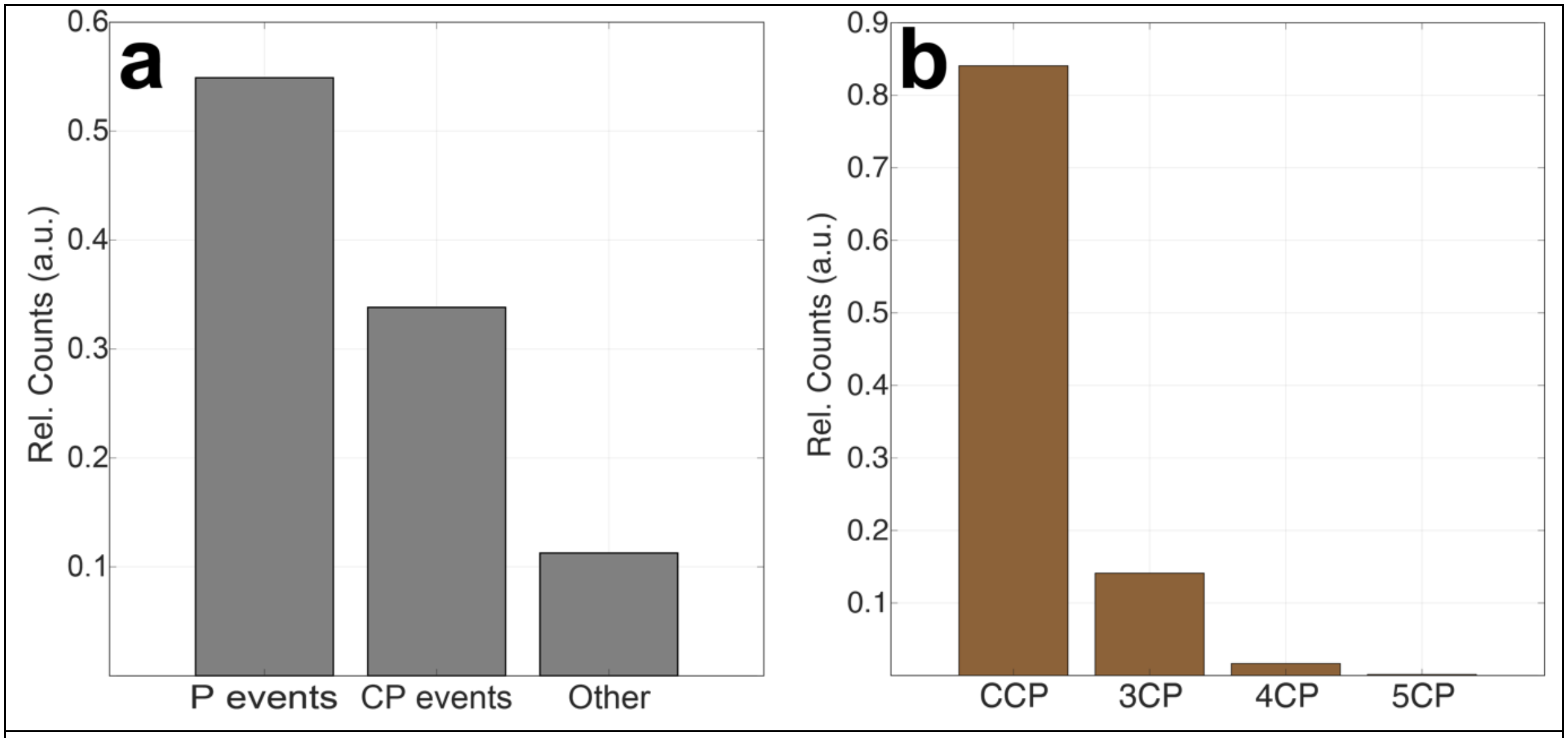

**Figure 5.** (**a**) Distribution of photoelectric (P), Compton followed by photoelectric (CP) and multi-scatter ('other') events based on the simulation of 511-keV gamma rays impinging all crystal arrays in the scanner. (**b**) Distribution of multi-scatter events within the scanner.

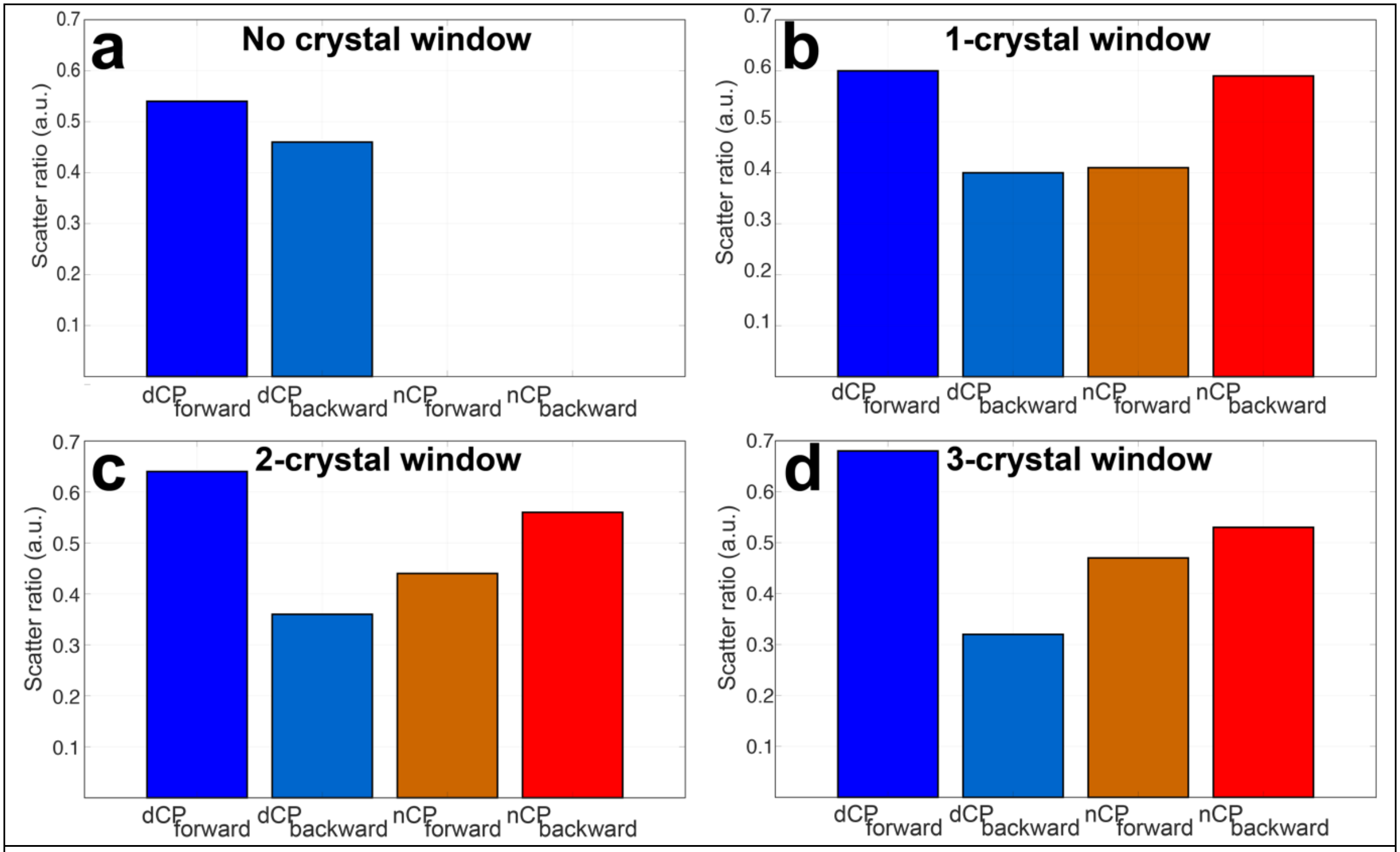

**Figure 6.** (**a**) Forward and backward scatter ratios for dCP events without a crystal window. (**b-d**) Forward and backward scatter ratios for nCP and dCP events while varying the crystal window from 1 to 3 crystals, respectively.

## 3.2. Network Performance

### 3.2.1. Event Classification Accuracy

Figure 7 shows the confusion matrices for DNN-based event classification accuracy for the three point sources. In all cases, the DNN exhibited 90% accuracy in classifying

P events. For nCP and dCP events, the accuracy was consistently ≥ 70% and ≥ 62%, respectively. The most common classification error for nCP events was misclassifying them as P events; for dCP events, the most common classification error was misclassifying them as nCP events. 11-13% of dCP events were erroneously classified as photoelectric events. Inspection of such events showed that 53% of these corresponded to events in which either the energy deposition during the first interaction was ≤ 100 keV or an edge/corner crystal was involved. Overall, the classification accuracy for all event types across all point source locations was 82%. The classification accuracy of CP (ncP and dCP) events across all point source locations was 84%.

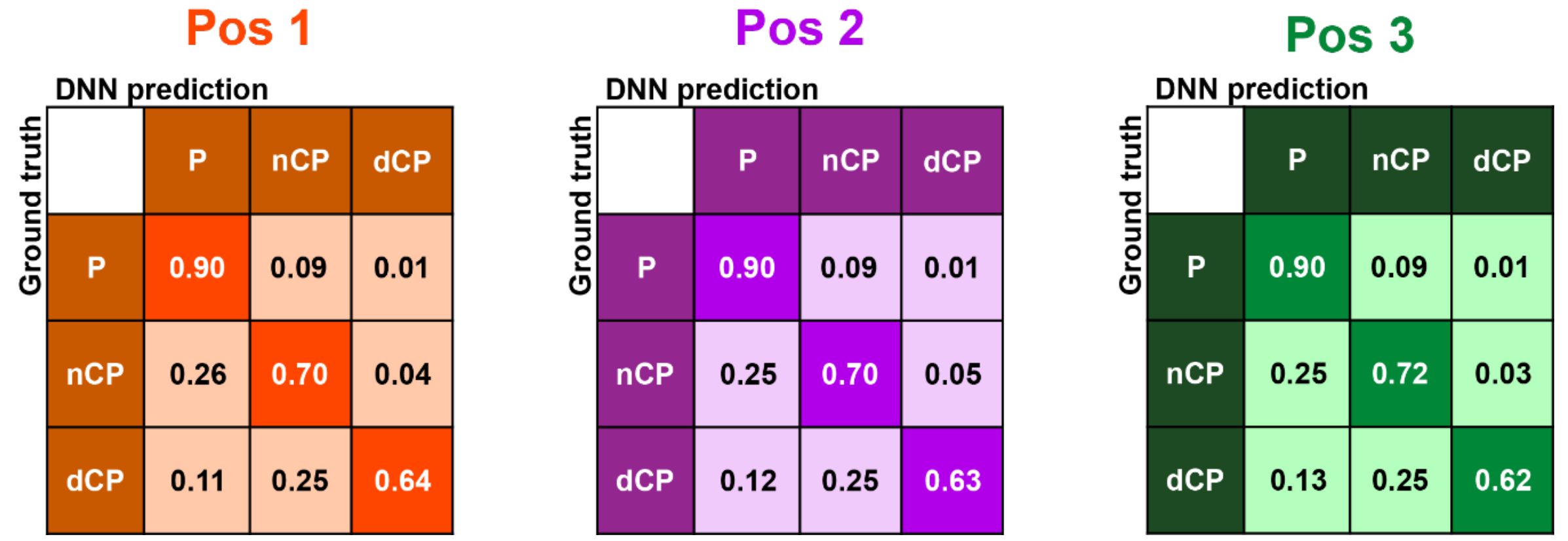

**Figure 7.** Confusion matrices of DNN classification accuracy for the three point sources, Pos 1 (orange), Pos 2 (purple) and Pos 3 (green), for the different event types (P, nCP and dCP).

3.2.2. Crystal Identification Accuracy

Figure 8 shows the *L1* distance between the true and DNN-predicted crystal indices involved in P events (top) and CP events (bottom). Since the network does not predict the crystal indices in order (i.e., $\boldsymbol{c}_a$ is not necessarily $\boldsymbol{c}_1$), we calculated the distance as the minimum distance between the two possible cases determined by the network:

$$\min(\|\boldsymbol{c}_1 - \boldsymbol{c}_a\|_1 + \|\boldsymbol{c}_2 - \boldsymbol{c}_b\|_1, \|\boldsymbol{c}_1 - \boldsymbol{c}_b\|_1 + \|\boldsymbol{c}_2 - \boldsymbol{c}_a\|_1) \tag{6}$$

Table 1 shows the mean *L1* distance between the true and DNN-predicted crystal indices for the different point source locations and event types. The mean distance for all event types (P + CP events) is given as a metric for the overall network performance.

3.2.3. CP Energy Prediction

Figure 9 illustrates the disparity between the actual energy deposited in a crystal during the first interaction of a CP event and the predicted energy after applying forward or backward scatter assumptions (Section 2.3.1). Although the energy prediction is relatively poor, it does function to guide the network in estimating the first crystal of interaction in CP events, as will be shown in section 3.2.4. Across all three point source positions, 43% of events fell within this ±50 keV range. 66% of events were within 100 keV of the true energy.

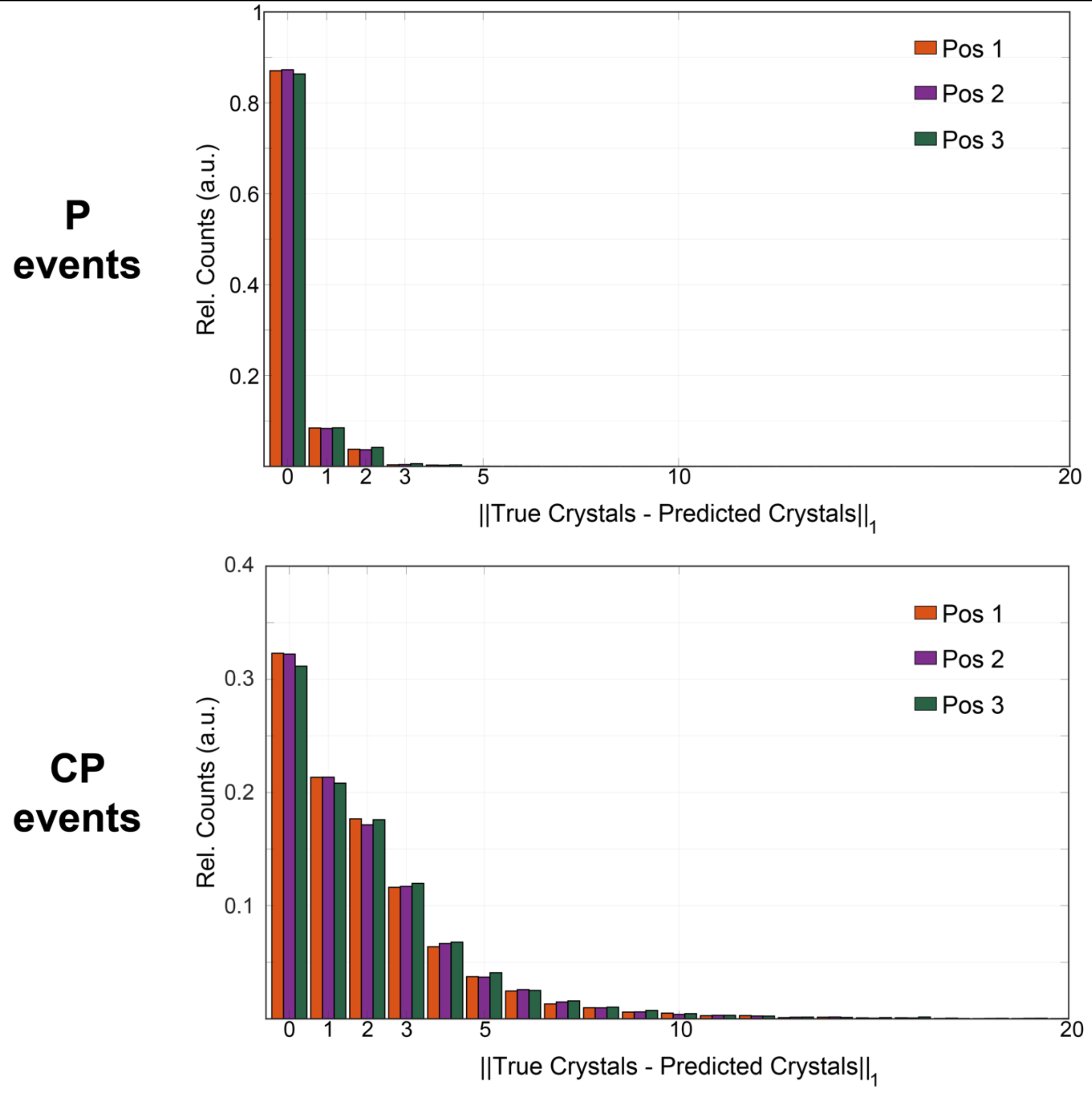

**Figure 8**. *L1* distance between the true (simulation) and predicted (DNN) crystal indices involved in P (**top**) and CP (**bottom**) events from the three different points source locations: Pos 1, orange; Pos 2, purple; Pos 3, green.

Table 1. Mean *L1* distance values between true and DNN-predicted crystals for P, CP and all events (P + CP) across the different point source locations

| Source position | **P events** | **CP events** | **All events** |
|---|---|---|---|
| Pos 1 | 0.19 | 1.95 | 0.79 |
| Pos 2 | 0.18 | 1.97 | 0.79 |
| Pos 3 | 0.20 | 2.04 | 0.83 |

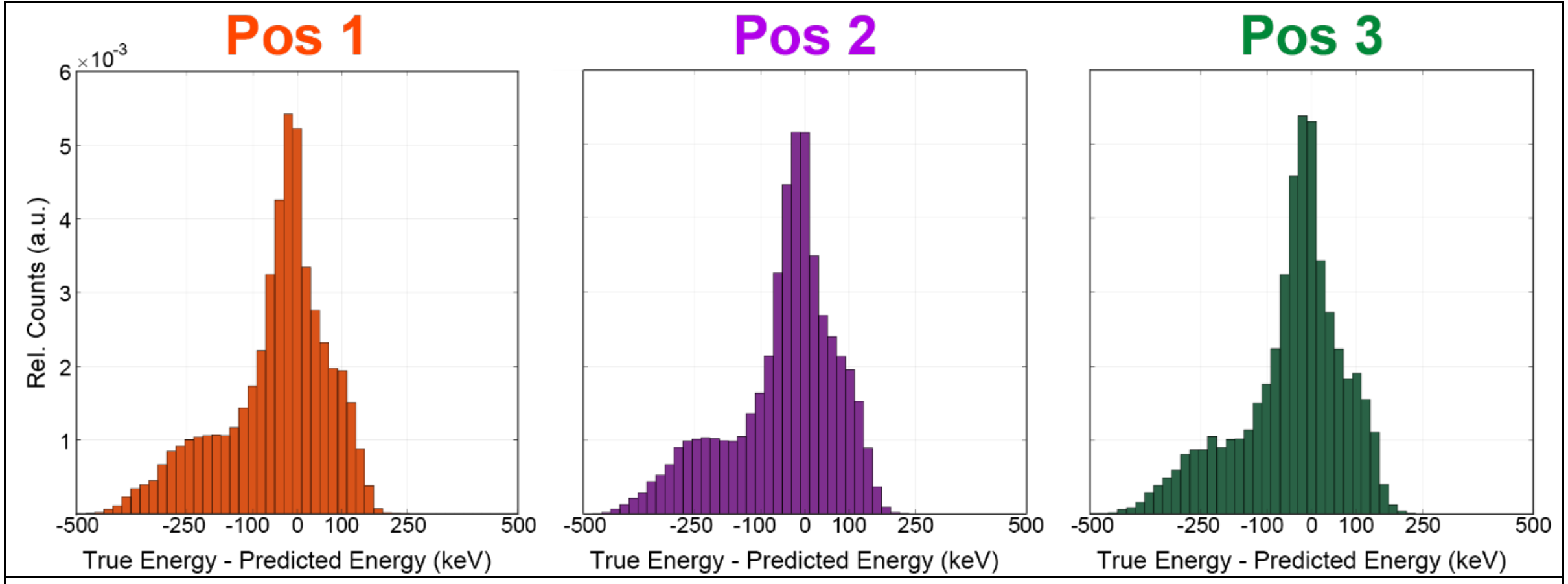

**Figure 9**. Energy difference between the true and predicted energy deposition during the scattering of CP events. From left to right, the energy differences for events from Pos 1 (orange), Pos 2 (purple) and Pos 3 (green) sources.

3.2.4. Positioning Performance

Figure 10 compares the DNN-based and Anger logic-based positioning performance for the three point source locations for both P events and CP events. Accuracy is reported for two scenarios: (i) accepting events only when the crystal difference was zero, and (ii) allowing a 1-crystal window error. For case 1 (exact), the DNN and Anger based approaches achieved an accuracy of 92-93% and 55-58%, respectively, for P events. For CP events, the accuracy was 26-27% (DNN) and 11-12% (Anger logic). Considering all events for case 1, the positioning accuracy was 70% (DNN) and 42% (Anger logic). For case 2 (1-crystal margin) and considering all events, the accuracy was 87% (DNN) and 82% (Anger logic).

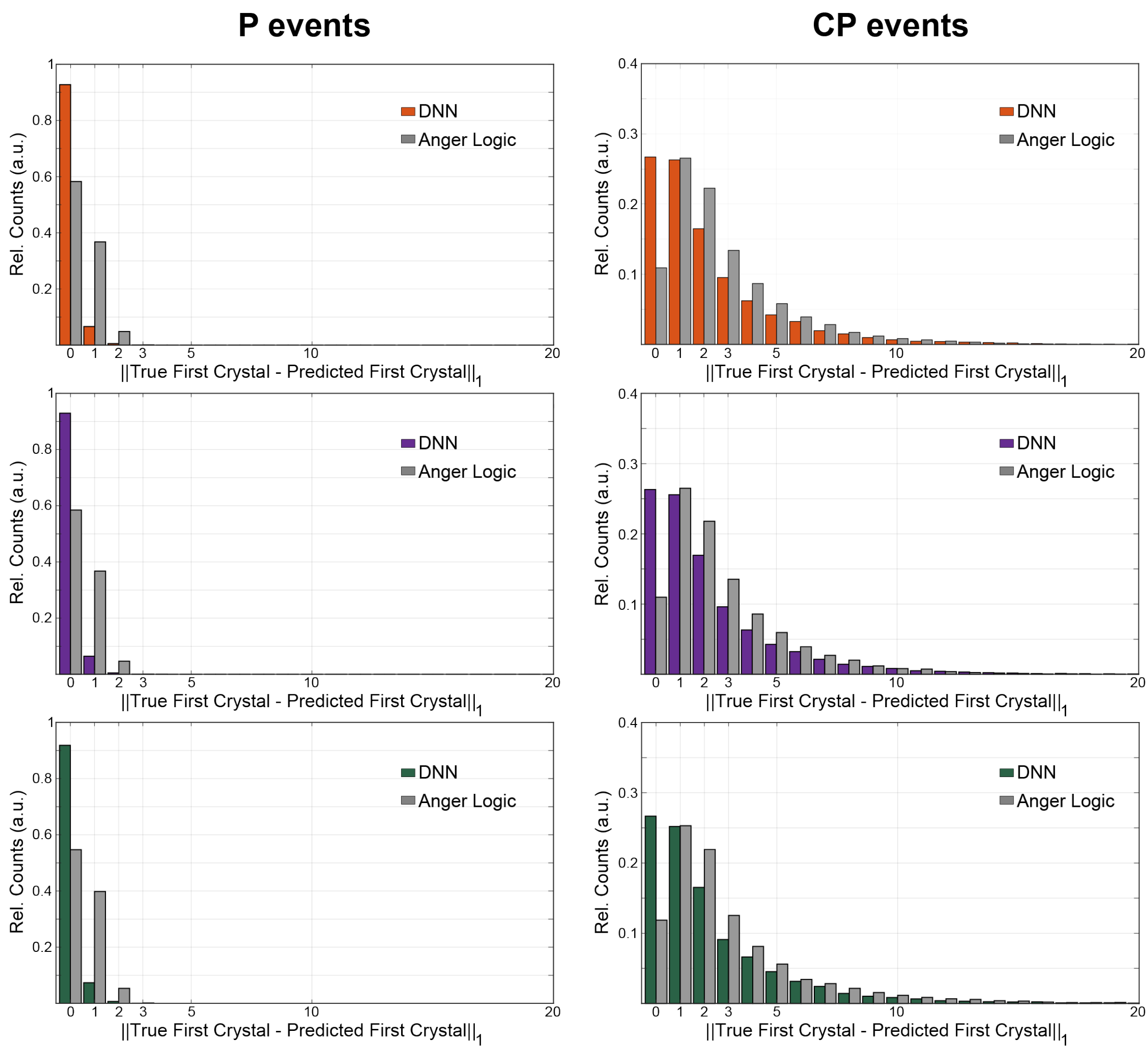

**Figure 10.** Positioning performance comparison between the DNN- and Anger logic-based approach for the three point sources: Pos 1 (**top**), Pos 2 (**middle**) and Pos 3 (**bottom**). Positioning comparison for P events (**Left**) and CP events (**Right**). In all cases, the coloured bars indicate the DNN performance and the light grey bars indicate Anger logic performance.

## Discussion

### 4.1. ICS Forward and Backward Scatter Ratios

The limited time resolution in PET detectors due to scintillation processes and readout electronics complicates the use of pulse discrimination methods in light-sharing detectors for distinguishing multiple scatter interactions in ICS events. Instead, ICS event discrimination most commonly relies on identifying atypical light distribution patterns on the detector, which differ from those generated by single photoelectric interactions (Lee and Lee 2021, 2023, Petersen et al 2024). However, once these atypical patterns are recognized, questions arise: which two crystals (in the case of CP events) are responsible for these patterns, and in which crystal did the scatter interaction occur? These questions do not have straightforward answers. Assuming knowledge of the two interacting crystals and their respective energy depositions, one might naturally apply the Klein-Nishina formulation (Klein and Nishina 1929) to select the crystal with the lowest energy deposition as the first crystal. The results in section 3.1 demonstrate that, for 20-mm thick crystal needles (width < 1 mm), such an assumption is poor. The scatter ratios are nearly equal for forward and backward scattering, and which results from the different photoelectric attenuation lengths ($\lambda_p$) of the scattered photons (Petersen et al 2024). With such constraint, we implemented crystal windows to determine the most appropriate ratio for categorizing events as forward or backward scattered. We found that a 1-crystal window results in forward-backward ratios of 60-40 and 40-60 for dCP and nCP events, respectively. However, these ratios imply that, even in an ideal scenario where our network achieved 100% accuracy in crystal and energy predictions, the positioning method would be penalized 40% of the time due to the inherent physics of the problem. Expanding the crystal window to more than one crystal introduces the drawback of nearly equalizing the ratios for nCP events in terms of forward and backward scattering, rendering wider windows impractical.

## 4.2. Network Performance

The network distinguished photoelectric events and single-scatter events with an accuracy exceeding 80%. This result is comparable with previously reported performance (Lee and Lee 2021), however in our study this was obtained using multiplexed photodetector signals rather than the individual signals. Network performance identifying the 2 crystals involved in CP events was ~32%. This relatively low value can be attributed to the non-uniqueness of decomposing CP light distributions in terms of two energy-weighted basis distributions can lead to different combinations of crystal basis distributions and energy weights giving the same or very similar CP row-column light distributions (see Fig. 11a). The latter is particularly true for basis distributions of crystals adjacent to the actual impinged crystals.

For P events, where energy prediction does not guide crystal positioning, the network output values ranging between 0 and 1 do not pose an issue. However, for CP events, the energy prediction accuracy of the network is penalised by the forward and backward energy assumptions. Figure 11b presents the network energy performance for CP events without these assumptions. In this case, we used the approach followed in section 3.2.2 (equation 6). Once one of the two cases was chosen, the energy difference between the true first energy ($e_1$) and the energy from the crystal subtracting $\boldsymbol{c}_1$ was calculated. For example, for a case in which $\|\boldsymbol{c}_1 - \boldsymbol{c}_a\|_1 + \|\boldsymbol{c}_2 - \boldsymbol{c}_b\|_1 < \|\boldsymbol{c}_1 - \boldsymbol{c}_b\|_1 + \|\boldsymbol{c}_2 - \boldsymbol{c}_a\|_1$, an 'unbiased' estimation of the network energy performance is $e_1 - e_a$. Within a ±50 keV energy window and without scatter assumptions, the network energy prediction accuracy is measured to be 12% higher than the accuracy obtained after scatter assumptions (section 3.2.3).

Crystal identification performance for P events by the network is notably higher compared to Anger logic (87% v 53%) primarily because of better performance on edge crystals. For CP events, the DNN algorithm improves the crystal positioning by 15% with respect to Anger logic. Positioning of CP events using Anger logic shows that, across all point sources, a one crystal difference is most likely to be observed (Figure 10). This systematic error is likely to arise from the shifting of the centre of gravity of the detected scintillation light due to the double interaction of the gamma ray within the crystal.

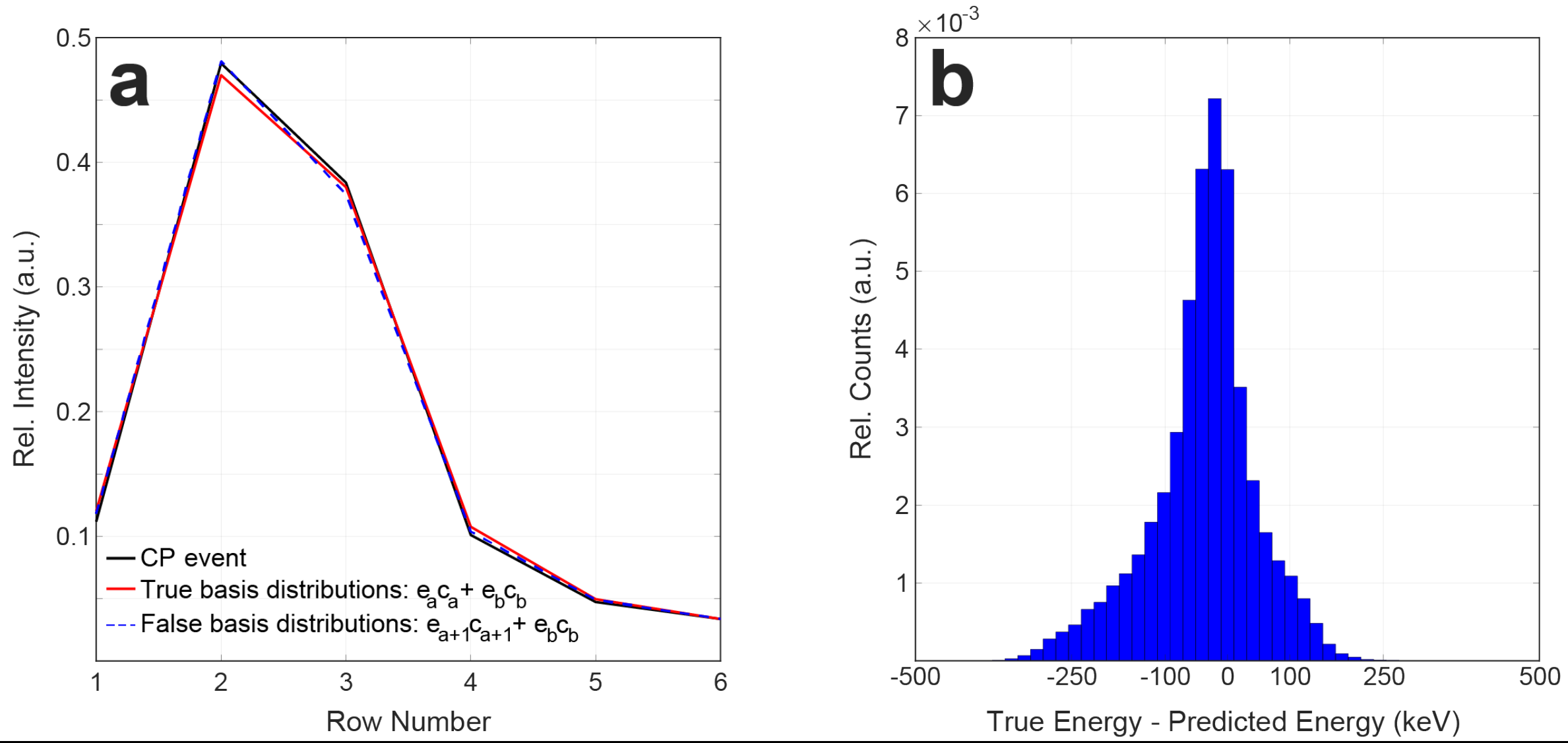

**Figure 11**. (**a**) Solid black line: a CP-event row light distribution from GATE simulations (section 2.3) involving crystals $\boldsymbol{c}_a$ and $\boldsymbol{c}_b$; solid red line: energy-weighted combination of the row light distributions of true two crystals involved in the CP event; dashed blue line: a row light distribution obtained using the wrong basis distributions ($\boldsymbol{c}_{a+1}$ instead of $\boldsymbol{c}_a$) and wrong energy weights ($e_{a+1} \neq e_a$). (**b**) Unbiased energy prediction performance of the DNN (i.e., without scatter assumptions).

### 4.3. 3D Positioning (DOI)

Modern PET scanners now integrate DOI to mitigate parallax errors. Although it was beyond the scope of this study, incorporating DOI into the DNN-based event positioning method could be achieved via at least two adaptations:

(i) DOI could be calculated as a post-processing step from the DNN results. For example, after obtaining an initial DOI estimate based on the energy ratio between both photodetectors, an updated estimate could then be determined after classifying the event into nCP or dCP. The forward and backward scatter assumptions in both groups would then be applied to rebin the calculated DOI value into its 'true' bin.

(ii) A more ambitious approach is to incorporate DOI directly into the network predictions (as in Petersen et al 2024). However, this would necessitate modifications to the current proof-of-concept method. Specifically, the network inputs should transition from a row and column light distribution to a 24-entry vector, representing the row and column light distributions from each photodetector. An alternative normalisation technique preserving relative differences between photodetectors would be required.

In this work, we restricted our focus to a 2D DNN-based positioning algorithm. Our findings, based on Monte Carlo simulated data, indicate that this algorithm, with an event positioning accuracy of 70%, outperforms the conventional Anger logic method (42% accuracy) by 28%. Given that PET relies on the coincidence detection of two annihilation photons, the 2D coincidence positioning accuracy of the DNN algorithm is estimated to be 49%, surpassing the 18% achieved by Anger logic. Future work will extend this proof-of-concept algorithm to encompass DOI considerations, with implementation on a physical PET detector.

## Conclusions

We successfully developed a proof-of-concept 2D positioning algorithm utilising deep neural networks for light-sharing and multiplexed PET detectors. The algorithm can identify single scatter events from pure photoelectric events with an accuracy of 82%. For event positioning, the method demonstrates a 28% improvement over Anger logic for both photoelectric and single-scatter interactions. Since PET relies in the coincidence detection of two photons, the coincidence positioning accuracy of the DNN algorithm is expected to be ~2.7 times higher than that achieved with Anger logic.

## Acknowledgments

This work was supported by the Australian Research Council (Discovery grant DP190102318).

F.E. Enríquez-Mier-y-Terán was supported by the Mexican National Council of Humanities, Sciences and Technologies (CONAHCYT).

## References

Casey M E and Nutt R 1986 A multicrystal two dimensional BGO detector system for positron emission tomography *IEEE Trans Nucl Sci* **33** 460–3

Decuyper M, Stockhoff M, Vandenberghe S and Van Holen R 2021 Artificial neural networks for positioning of gamma interactions in monolithic PET detectors *Phys Med Biol* **66**

Enriquez-Mier-Y-Teran F E, Brandt O, Kwon S I, Bai X, Bec J, Judenhofer M S, Peng P, Cherry S R, Meikle S R and Kyme A Z 2021 Open-Field Mouse Brain PET: Towards a System for Simultaneous Brain PET and Behavioral Analysis in Small Animals *2021 IEEE Nuclear Science Symposium and Medical Imaging Conference Record, NSS/MIC 2021 and 28th International Symposium on Room-Temperature Semiconductor Detectors, RTSD 2022*

Gonzalez A J, Berr S S, Cañizares G, Gonzalez-Montoro A, Orero A, Correcher C, Rezaei A, Nuyts J, Sanchez F, Majewski S and Benlloch J M 2018 Feasibility study of a small animal PET insert based on a single LYSO monolithic tube *Front Med (Lausanne)* **5**

Gonzalez-Montoro A, Gonzalez A J, Pourashraf S, Miyaoka R S, Bruyndonckx P, Chinn G, Pierce L A and Levin C S 2021 Evolution of PET Detectors and Event Positioning Algorithms Using Monolithic Scintillation Crystals *IEEE Trans Radiat Plasma Med Sci* **5** 282–305

Iborra A, González A J, González-Montoro A, Bousse A and Visvikis D 2019 Ensemble of neural networks for 3D position estimation in monolithic PET detectors *Phys Med Biol* **64**

Jaliparthi G, Martone P F, Stolin A V and Raylman R R 2021 Deep residual-convolutional neural networks for event positioning in a monolithic annular PET scanner *Phys Med Biol* **66**

St. James S and Thompson C J 2006 Image blurring due to light-sharing in PET block detectors *Med Phys* **33** 405–10

Kang H G, Nishikido F and Yamaya T 2021 A staggered 3-layer DOI PET detector using BaSO4 reflector for enhanced crystal identification and inter-crystal scattering event discrimination capability *Biomed Phys Eng Express* **7**

Klein O and Nishina Y 1929 Über die Streuung von Strahlung durch freie Elektronen nach der neuen relativistischen Quantendynamik von Dirac *Zeitschrift für Physik* **52** 853–68

Kyme A Z, Judenhofer M S, Gong K, Bec J, Selfridge A, Du J, Qi J, Cherry S R and Meikle S R 2017 Open-field mouse brain PET: Design optimisation and detector characterisation *Phys Med Biol* **62** 6207–25

Van Der Laan D J, Schaart D R, Maas M C, Beekman F J, Bruyndonckx P and Van Eijk C W E 2010 Optical simulation of monolithic scintillator detectors using GATE/GEANT4 *Phys Med Biol* **55** 1659–75

Labella A, Vaska P, Zhao W and Goldan A H 2020 Convolutional Neural Network for Crystal Identification and Gamma Ray Localization in PET *IEEE Trans Radiat Plasma Med Sci* **4** 461–9

Lee M S, Kang S K and Lee J S 2018 Novel inter-crystal scattering event identification method for PET detectors *Phys Med Biol* **63**

Lee S and Lee J S 2023 Experimental evaluation of convolutional neural network-based inter-crystal scattering recovery for high-resolution PET detectors *Phys Med Biol* **68**

Lee S and Lee J S 2021 Inter-crystal scattering recovery of light-sharing PET detectors using convolutional neural networks *Phys Med Biol* **66**

Levin A and Moisan C 1996 A More Physical Approach to Model the Surface Treatment of Scintillation Counters and its Implementation into DETECT *IEEE Nuclear Science Symposium. Conference Record* pp 702–6

Liu Z, Mungai S, Niu M, Kuang Z, Ren N, Wang X, Sang Z and Yang Y 2023 Edge effect reduction of high-resolution PET detectors using LYSO and GAGG phoswich crystals *Phys Med Biol* **68**

Petersen E, LaBella A, Li Y, Wang Z and Goldan A H 2024 Resolving inter-crystal scatter in a light-sharing depth-encoding PET detector *Phys Med Biol* **69**

Pichler B J, Swann B K, Rochelle J, Nutt R E, Cherry S R and Siegel S B 2004 Lutetium oxyorthosilicate block detector readout by avalanche photodiode arrays for high resolution animal PET *Phys Med Biol* **49** 4305–19

Roncali E, Stockhoff M and Cherry S R 2017 An integrated model of scintillator-reflector properties for advanced simulations of optical transport *Phys Med Biol* **62** 4811–30

Sanaat A and Zaidi H 2020 Depth of interaction estimation in a preclinical PET scanner equipped with monolithic crystals coupled to SiPMs using a deep neural network *Applied Sciences (Switzerland)* **10**

Sarrut D, Bała M, Bardi s M, Bert J, Chauvin M, Chatzipapas K, Dupont M, Etxebeste A, Fanchon L M, Jan S, Kayal G, Kirov A S, Kowalski P, Krzemien W, Labour J, Lenz M, Loudos G, Mehadji B, Ménard L, Morel C, Papadimitroulas P, Rafecas M, Salvadori J, Seiter D, Stockhoff M, Testa E, Trigila C, Pietrzyk U, Vandenberghe S, Verdier M A, Visvikis D, Ziemons K, Zvolský M and Roncali E 2021 Advanced Monte Carlo simulations of emission tomography imaging systems with GATE *Phys Med Biol* **66**

Shao Y, Cherry S R, Siegel S and Silverman R W 1996 *A Study of Inter-Crystal Scatter in Small Scintillator Arrays Designed for High Resolution PET Imaging* vol 43

Shinohara K, Suga M, Yoshida E, Nishikido F, Inadama N, Tashima H and Yamaya T 2014 Maximum likelihood estimation of inter-crystal events scattering events for light sharing PET detectors *2014 IEEE Nuclear Science Symposium and Medical Imaging Conference (NSS/MIC)* (Seattle, WA, USA) pp 1–3

Song T Y, Wu H, Komarov S, Siegel S B and Tai Y C 2010 A sub-millimeter resolution PET detector module using a multi-pixel photon counter array *Phys Med Biol* **55** 2573–87

Stickel J R, Qi J and Cherry S R 2007 *Fabrication and Characterization of a 0.5-mm Lutetium Oxyorthosilicate Detector Array for High-Resolution PET Applications* vol 48

Stockhoff M, Van Holen R and Vandenberghe S 2019 Optical simulation study on the spatial resolution of a thick monolithic PET detector *Phys Med Biol* **64**

Stockhoff M, Jan S, Dubois A, Cherry S R and Roncali E 2017 Advanced optical simulation of scintillation detectors in GATE V8.0: First implementation of a reflectance model based on measured data *Phys Med Biol* **62** L1–8

Vincent L and Soille P 1991 Watersheds in digital spaces: an efficient algorithm based on immersion simulations *IEEE Trans Pattern Anal Mach Intell* **13** 583–98

Yamamoto S, Watabe H, Watabe T, Ikeda H, Kanai Y, Ogata Y, Kato K and Hatazawa J 2016 Development of ultrahigh resolution Si-PM-based PET system using 0.32 mm pixel scintillators *Nucl Instrum Methods Phys Res A* **836** 7–12

Yang Q, Kuang Z, Sang Z, Yang Y and Du J 2019 Performance comparison of two signal multiplexing readouts for SiPM-based pet detector *Phys Med Biol* **64**

Yang Y, Bec J, Zhou J, Zhang M, Judenhofer M S, Bai X, Di K, Wu Y, Rodriguez M, Dokhale P, Shah K S, Farrell R, Qi J and Cherry S R 2016 A prototype high-resolution Small-Animal PET scanner dedicated to mouse brain imaging *Journal of Nuclear Medicine* **57** 1130–5

Zhang C, Sang Z, Wang X, Zhang X and Yang Y 2019 The effects of inter-crystal scattering events on the performance of PET detectors *Phys Med Biol* **64**